\RequirePackage{ifpdf}
\ifpdf 
\documentclass[pdftex]{sigma}
\else
\documentclass{sigma}
\fi

\def\ov{\overline}
\def\a{\alpha}
\def\b{\beta}

\def\l{\lambda}
\def\L{\Lambda}

\def\t{\theta}

\def\wt{\widetilde}
\def\BC{{\mathbb C}} 
\def\BR{{\mathbb R}} 

\numberwithin{equation}{section}

\begin{document}

\allowdisplaybreaks

\renewcommand{\PaperNumber}{054}

\FirstPageHeading

\renewcommand{\thefootnote}{$\star$}

\ShortArticleName{Non-Isospectral Canonical System}

\ArticleName{B\"acklund--Darboux Transformation\\ for
Non-Isospectral Canonical System\\ and Riemann--Hilbert
Problem\footnote{This paper is a contribution to the Vadim
Kuznetsov Memorial Issue `Integrable Systems and Related Topics'.
The full collection is available at
\href{http://www.emis.de/journals/SIGMA/kuznetsov.html}{http://www.emis.de/journals/SIGMA/kuznetsov.html}}}

\Author{Alexander SAKHNOVICH} \AuthorNameForHeading{A. Sakhnovich}

\Address{Fakult\"at f\"ur Mathematik, Universit\"at Wien,
Nordbergstrasse 15, A-1090 Wien, Austria}
\Email{\href{mailto:al_sakhnov@yahoo.com}{al\_sakhnov@yahoo.com}}

\ArticleDates{Received October 25, 2006, in f\/inal form March 19,
2007; Published online March 25, 2007}

\Abstract{A GBDT version of the  B\"acklund--Darboux
transformation is constructed for a~non-isospectral canonical
system, which plays essential role in the theory of random matrix
models. The corresponding Riemann--Hilbert problem is treated and
some explicit formulas are obtained. A related inverse problem is
formulated and solved.}


\Classification{35Q15; 37K35}

\section{Introduction}

We shall consider  a non-isospectral  system
\begin{gather} \label{1.1}
w_x(x,z)=i \l J H(x) w(x, z), \qquad \l=(z-x)^{-1},
\end{gather}
where $w_x= \frac{d}{d x}w$, and $J$ and $H(x)$ are $m \times m$
matrices:
\begin{gather*} 
H(x)=H(x)^*, \qquad J=J^*=J^{-1}.
\end{gather*}
When the Hamiltonian $H \geq 0$, and the spectral parameter $\l$
does not depend on $x$, the system above is a classical canonical
system. A version of the B\"acklund--Darboux transformation (BDT)
for the classical canonical system have been constructed in
\cite{SaA3}. In our case (\ref{1.1}) the spectral parameter
$\l=(z-x)^{-1}$ depends on $x$, and here we construct BDT for
this case.

BDT is a fruitful approach to obtain solutions of the linear
dif\/ferential equations and systems. It is also widely used to
construct explicit solutions of  integrable nonlinear systems. For
that purpose BDT is applied simultaneously to two auxiliary linear
systems of the integrable one. BDT is closely related to the
symmetry properties. Since the original works of B\"acklund and
Darboux, a much deeper understanding of this transformation has
been achieved and various   interesting versions of the
B\"acklund--Darboux transformation have been introduced (see, for
instance,  \cite{C, Dei0, Ge,
 Gu, Mar, MS, Mi,  T, ZM}). Important works on the
B\"acklund--Darboux transformation both in the continuous and
discrete cases have been written by V.B. Kusnetzov and his
coauthors (see \cite{KPR, KSS} and references therein).

We apply BDT to construct explicitly new solutions of the
Riemann--Hilbert problem on the interval $[0, \, l]$:
\begin{gather} \label{1.24}
W_+(s)=W_-(s)R(s)^2, \qquad 0 \leq s \leq a,
\end{gather}
where $W(z)$ is analytic for $z \notin [0, \, a]$, and $W(z) \to
I_m$, when $z \to \infty$, $I_m$ is the $m \times m$ identity
matrix. For important classes of $R$ the solution of  problem
(\ref{1.24})  takes the form
\begin{gather} \label{1.24'}
W(z)=w(l,z), \qquad W_+(s)= \lim_{\eta \to +0} w(l, s+i \eta),
\qquad W_-(s)= \lim_{\eta \to +0} w(l, s-i \eta),
\end{gather}
where the $m \times m$ fundamental solution $w$ of (\ref{1.1}) is
normalized by the condition
\begin{gather} \label{1.2'}
w(0,z)=I_m.
\end{gather}
The necessary and suf\/f\/icient conditions for (\ref{1.24'}) are
given in \cite[p.~209]{SaL4} (see also \cite{SaL5, SaL3}). It is
useful to obtain explicit formulas for $H$ and $R$.

The problem (\ref{1.24})  is of interest in the random matrix
theory: the Markov parameters appearing in the series
representation $w(l, z)=I_m+z^{-1}M_1(l)+z^{-2}M_2(l)+\cdots$ are
essential for the random matrices problems \cite{Dei00, Dei}. In
particular, in the bulk scaling limit of the Gaussian unitary
ensemble of Hermitian matrices the probability that an interval of
length $l$ contains no eigenva\-lues is given by the function
$P(l)$, which satisf\/ies the equality $ \frac{d}{dl}\log \,
P(l)=i\big(m_{22}(l)-m_{11}(l)\big)$, where $m_{11}$ and $m_{22}$
are the corresponding entries of the $2 \times 2$  matrix $
M_{1}$. Notice that $  M_1(l)=\int_0^lJH(x)dx$.

When $J=I_m$, system (\ref{1.24})  is essential in the prediction
theory \cite{W}.

We construct a B\"acklund--Darboux transformation  for system
(\ref{1.1}) in the next Section~2. Section 3 is dedicated to
explicit solutions, and Section~4 is dedicated to an inverse
problem.

\section[B\"acklund-Darboux transformation]{B\"acklund--Darboux transformation}\label{sec1}

To construct B\"acklund--Darboux transformation  we use the
methods developed in \cite{SaA2,SaA3} for non-isospectral problems
and canonical system, respectively. For this purpose f\/ix integer
$n>0$ and $n \times n$ parameter matrices $A(0)$, $S(0)=S(0)^*$.
Fix also $n \times m $ parameter matrix $\Pi(0)$ so that the
matrix identity
\begin{gather} \label{1.3}
A(0)S(0)-S(0)A(0)^*=i \Pi(0)J \Pi(0)^*
\end{gather}
holds. Introduce now matrix functions $A(x)$, $S(x)$ and $\Pi(x)$
by their values at $x=0$ and equations
\begin{gather} \label{1.4}
A_x=A^2, \qquad \Pi_x=- i A \Pi J H,
\\ \label{1.5}
S_x= \Pi J H J^* \Pi^*- (AS+SA^*).
\end{gather}
Then it can be checked by direct dif\/ferentiation that the matrix
identity
\begin{gather} \label{1.6}
AS-SA^*=i \Pi J \Pi^*
\end{gather}
holds for each $x$. Notice that the equation $A_x=A^2$ is
motivated by the similar equation $\l_x= \l^2$ for the spectral
parameter $\l$ because $A$ can be viewed as  a generalized
spectral parameter (see~\cite{SaA2}). In the points of
invertibility of $S$ we can introduce a transfer matrix function
in the Lev Sakhnovich form \cite{SaL1,SaL2,SaL3}
\begin{gather} \label{1.7}
w_A(x, z)=I_m-i J \Pi(x)^*S(x)^{-1}(A- \l I_n)^{-1} \Pi(x).
\end{gather}
This transfer matrix function has an important $J$-property
\cite{SaL1}:
\begin{gather} \label{1.7'}
w_A(x, \ov{z})^*Jw_A(x, z)=J, \qquad \mathrm{i.e.,} \qquad
w_A(x,z)^{-1}=Jw_A(x, \ov{z})^*J.
\end{gather}
Put
\begin{gather} \label{1.10}
 v(x, z)=w_0(x)^{-1}w_A(x,z),
\end{gather}
where matrix function $w_0$ is def\/ined by the relations
\begin{gather} \label{1.10'}
\frac{d}{dx}w_0(x)= \wt G_0(x)  w_0(x), \qquad w_0(0)^*J w_0(0)=J,
\\ \label{1.10''}
 \wt G_0=-J (i \Pi^*S^{-1} \Pi- H J \Pi^* S^{-1} \Pi+
\Pi^* S^{-1} \Pi J^* H)
\end{gather}
up to $J$-unitary initial value $w_0(0)$. (We omit sometimes
argument $x$ in the formulas for brevity.)
\begin{theorem} \label{TmBDT} Suppose $w$ is the
fundamental solution of system \eqref{1.1} and relations
\eqref{1.3}--\eqref{1.5} are valid. Then in the points of
invertibility of $S(x)$ the matrix function
\begin{gather} \label{1.14}
\wt w(x, z)=v(x, z) w(x, z)
\end{gather}
is well defined and satisfies the transformed system
\begin{gather} \label{1.13}
\frac{d}{dx} \wt w=i \l J \wt H \wt w,
\end{gather}
where
\begin{gather} \label{1.12}
\wt H(x)=w_0(x)^*H(x)w_0(x).
\end{gather}
Moreover, if $\det \, S(x) \, \not= \, 0$ $(0 \leq x \leq l)$ then
the fundamental solution of system \eqref{1.13} is given by the
formula
\begin{gather} \label{1.15}
\wt w(x, z)=v(x, z) w(x, z)v(0, z)^{-1}, \qquad 0 \leq x \leq l.
\end{gather}
\end{theorem}

\begin{proof}
The proof is based on the equation for the transfer matrix
function
\begin{gather} \label{1.9}
\frac{d}{d x} w_A(x, z)=\wt G(x, z)w_A(x, z)-i \l w_A(x, z) J
H(x),
\end{gather}
where
\begin{gather} \label{1.8}
\wt G(x, z)=i \l J H -J (i \Pi^*S^{-1} \Pi- H J \Pi^* S^{-1} \Pi+
\Pi^* S^{-1} \Pi J^* H).
\end{gather}
To prove (\ref{1.9}) consider f\/irst $\frac{d}{dx}J \Pi^*
S^{-1}$. By (\ref{1.4}) we have
\begin{gather} \label{1.16}
\frac{d}{dx}J \Pi^* S^{-1}= i J H J  \Pi^* A^*S^{-1}-J \Pi^*
S^{-1}S_xS^{-1}.
\end{gather}
Use now (\ref{1.5}) and (\ref{1.16}) to get
\begin{gather} \label{1.17}
\frac{d}{dx}J \Pi^* S^{-1}=(i J H J+J)\Pi^* A^*S^{-1}+ J \Pi^*
S^{-1}A- J \Pi^* S^{-1} \Pi J H J \Pi^* S^{-1}.
\end{gather}
Rewrite identity (\ref{1.6}) as $A^*S^{-1}=S^{-1}A-iS^{-1}\Pi J
\Pi^*S^{-1}$ and substitute this equality into (\ref{1.17}) to
obtain
\begin{gather}
\frac{d}{dx}J \Pi^* S^{-1}=(i J H J+2 J)\Pi^* S^{-1}A
\nonumber\\
 \label{1.18}
\phantom{\frac{d}{dx}J \Pi^* S^{-1}=}{}+J \big((H J-i I_m)
\Pi^*S^{-1} \Pi J- \Pi^* S^{-1} \Pi J H J \big) \Pi^* S^{-1}.
\end{gather}
We shall apply (\ref{1.18}) as well as the second relation in
 (\ref{1.4}) to dif\/ferentiate $w_A(x,z)$:
\begin{gather}
 \frac{d}{dx} w_A=-i \Big((i J H J+2 J)\Pi^*
S^{-1}\big((A- \l I_n) + \l I_n \big)(A- \l I_n)^{-1} \Pi
\nonumber\\
\phantom{\frac{d}{dx} w_A=}{}  + J \big((H J-i I_m) \Pi^*S^{-1}
\Pi J- \Pi^* S^{-1} \Pi J H J \big) \Pi^* S^{-1}(A- \l I_n)^{-1}
\Pi \Big)
\nonumber\\
\phantom{\frac{d}{dx} w_A=}{} +i J \Pi^* S^{-1}(A- \l
I_n)^{-1}(A_x - \l_x I_n) (A- \l I_n)^{-1} \Pi
\nonumber\\
\phantom{\frac{d}{dx} w_A=}{} -i J \Pi^* S^{-1}(A- \l I_n)^{-1}
\big(-i \big((A- \l I_n)+ \l I_n \big) \Pi J H \big).\label{1.19}
\end{gather}
Use substitutions $(A- \l I_n)^{-1}(A- \l I_n)=I_n$ and
\begin{gather*} 
(A- \l I_n)^{-1}(A_x - \l_x I_n) (A- \l I_n)^{-1}=I_n+2 \l (A- \l
I_n)^{-1},
\end{gather*}
and collect terms to rewrite (\ref{1.19}) in the form (\ref{1.9}).

According to formulas (\ref{1.10})--(\ref{1.10''}), (\ref{1.9}),
and (\ref{1.8}) we have
\begin{gather}
\frac{d}{d x}v(x,z)=w_0(x)^{-1}\big(\wt G(x,z) - \wt G_0(x) \big)
w_A(x,z)-i \l v(x,z)J H(x)\nonumber\\
 \label{1.21}
 \phantom{\frac{d}{d x}v(x,z)}{}
=i \l w_0(x)^{-1}J H(x)w_0(x)v(x,z)-i \l v(x,z)J H(x).
\end{gather}
Taking into account (\ref{1.10'}) we get
\begin{gather} \label{1.22}
w_0(x)^*J w_0(x)=w_0(0)^*J w_0(0)=J.
\end{gather}
Thus we rewrite (\ref{1.21}) as
\begin{gather} \label{1.11}
\frac{d}{d x} v(x, z)=i \l J \wt H(x)v(x, z)-i \l v(x, z) J H(x),
\end{gather}
where $\wt H$ is given by (\ref{1.12}). From (\ref{1.1}) and
(\ref{1.11}) it follows that (\ref{1.13}) is true for $\wt w$ of
the form~(\ref{1.14}).  In view of (\ref{1.2'}) one can see that
normalization (\ref{1.15}) yields $\wt w(0,z)=I_m$.
\end{proof}
Our next proposition provides conditions for invertibility of $S$.
\begin{proposition} \label{PnInvert}
Suppose matrix functions $H(x) \geq 0$ and $A(x)$ are summable on
the interval $[0, \, l]$, and $S(0)>0$. Then $S(x)>0$ for $0 \leq
x \leq l$, and so $S(x)$ is invertible.
\end{proposition}

\begin{proof} Put
\begin{gather} \label{M1}
Q(x)=V(x)S(x)V(x)^*, \qquad {\mathrm{where}} \qquad V_x=VA, \qquad
V(0)=I_n.
\end{gather}
Then in view of (\ref{1.5}) and (\ref{M1}) we have
\[
Q_x=V(S_x+AS+SA^*)V^*=V \Pi J H J^* \Pi^* V^* \geq 0, \qquad
Q(0)=S(0).
\]
It follows that
\begin{gather*}
Q(x)>0, \qquad S(x)=V(x)^{-1}Q(x) \big(V(x)^*
\big)^{-1}>0.\tag*{\qed}
\end{gather*}
\renewcommand{\qed}{}
\end{proof}

In view of the f\/irst equality in (\ref{1.4}) invertible matrix
function $A$ is of the form $A=(B-x I_n)^{-1}$. Further we shall
suppose that $A$ is def\/ined and both $A$  and $S$ are invertible
on some  interval $[0,l]$.
\begin{remark}
Suppose $A(x)$ and $S(x)$ are invertible on the interval $[0,l]$.
Using (\ref{1.17}) we can dif\/ferentiate
\begin{gather*} 
w_A(x, \infty):=I_m-i J \Pi^*S^{-1}A^{-1} \Pi.
\end{gather*}
In this way similarly to (\ref{1.9}) we can show that the matrix
function $w_0$, which satisf\/ies (\ref{1.10'}), (\ref{1.10''})
and initial condition $w_0(0)=U$, admits representation
\begin{gather} \label{yv1'}
w_0(x)=w_A(x, \infty) \wt U, \qquad \wt U=w_A(0, \infty)^{-1}U,
\qquad 0 \leq x \leq l.
\end{gather}
Notice that by (\ref{1.12}) and (\ref{1.22}) the equality $H J H
\equiv 0$ yields $\wt H J \wt H \equiv 0$, i.e., if $J H$ is
nilpotent, then $J \wt H$ is nilpotent too.
\end{remark}

Let now $w(l,z)$ satisfy Riemann--Hilbert equation (\ref{1.24})
where $W_{\pm}(s)= \lim\limits_{\eta \to +0}w(l,s \pm i \eta)$.
Suppose all conditions of Theorem \ref{TmBDT} are fulf\/illed.
Then, putting
\begin{gather*} 
\wt W_{\pm}(s)= \lim_{\eta \to +0} \wt w(l,s \pm i \eta),
\end{gather*}
we get
\begin{gather} \label{1.26}
\wt W_{\pm}(s)=v(l,s) W_{\pm}(s) v(0,s)^{-1}.
\end{gather}
In view of (\ref{1.24}) and (\ref{1.26}) we obtain
\begin{gather*} 
\wt W_+(s)= \wt W_-(s) v(0,s)R(s)^2 v(0,s)^{-1}= \wt W_-(s) \wt
R(s)^2,
\end{gather*}
where
\begin{gather} \label{1.28}
 \wt R(s)= v(0,s)R(s) v(0,s)^{-1}.
\end{gather}
The subcase of nilpotent matrix function $\wt R(s)-I_m$ is
important \cite{SaL4}. According to (\ref{1.28}) we have
\begin{gather} \label{1.28'}
 \wt R(s)-I_m= v(0,s) \big(R(s)-I_m \big) v(0,s)^{-1}.
\end{gather}
Hence, we get a corollary.
\begin{corollary}
 If $R(s)-I_m$ is
nilpotent, then $\wt R(s)-I_m$ is nilpotent too.
\end{corollary}
\section{Explicit solutions}\label{sec2}

 If we
know $A$, $S$, $\Pi$, then using the results of the previous
section we can construct explicit expressions for $\wt H$ and $\wt
R$. Consider the simplest case
\begin{gather} \label{yv2}
m=2, \qquad H=\b^* \b, \qquad \b \equiv [1 \quad i], \qquad 0 \leq
x \leq l, \qquad J=\left[\begin{array}{lr}0 & 1 \\ 1 &
0\end{array} \right].
\end{gather}
Then in formula (\ref{1.24}) we have
\begin{gather} \label{yv2'}
R(s) \equiv I_2+ \pi J \b^* \b, \qquad 0 < x < l.
\end{gather}
Indeed, in view of (\ref{1.1}) and (\ref{yv2}) we get
\begin{gather}
\b w_x(x,z)=0, \qquad \b J w_x(x,z)=2i(z-x)^{-1} \b w(x,z), \qquad
{\mathrm{i.e.,}}\nonumber
\\ \label{yvH1}
\b w(x,z)= \b w(0,z)= \b, \qquad \b J w(x,z)=-2 i \big( \ln \,
(z-x) \big) \b +{\mathrm{const}},
\end{gather}
where const means some constant (vector). In the f\/irst relation
in (\ref{yvH1}) we use normalization condition (\ref{1.2'}).
Taking into account (\ref{1.2'}) again, from the second relation
in (\ref{yvH1}) we derive
\begin{gather} \label{yvH2}
 \b J w(x,z)=2 i \left( \ln
\frac{z}{z-x} \right) \b +\b J.
\end{gather}
Equalities (\ref{yvH1}) and (\ref{yvH2}) imply that
\begin{gather}
\b W_+(s)= \b W_-(s)= \b, \qquad \b J W_+(s)=2 i \left( \left(\ln
\left| \frac{s}{s-l} \right| \right)- i \pi \right) \b +\b J,
\nonumber\\
 \label{yvH3}
\b J W_-(s)=2 i \left( \left(\ln \left| \frac{s}{s-l} \right|
\right)+ i \pi \right) \b +\b J.
\end{gather}
From (\ref{yvH3}) it follows that
\begin{gather} \label{yvH4}
TW_+(s)- T W_-(s)= \left[ \begin{array}{c} 0  \\ 4 \pi \b
\end{array} \right], \qquad T:=\left[ \begin{array}{c}
\b  \\  \b J
\end{array} \right].
\end{gather}
Notice that
\begin{gather} \label{yvH5}
T J T^*=2 J.
\end{gather}
So according to (\ref{yvH4})  we have
\begin{gather} \label{yvH6}
W_+(s)-  W_-(s)= \frac{1}{2}J T^* J\left[
\begin{array}{c} 0  \\ 4
\pi \b
\end{array} \right]=2 \pi J \b^* \b.
\end{gather}
Moreover, formula (\ref{yvH3}) implies that
\[
W_-(s)= \frac{1}{2}J T^* J\left[ \begin{array}{c} \b
\vspace{1mm}\\ \displaystyle 2 i \left( \left(\ln  \left|
\frac{s}{s-l} \right| \right)+ i \pi \right) \b +\b J
\end{array} \right].
\]
Hence, we obtain
\begin{gather} \label{yvH7}
J \b^*=W_-(s)J \b^*.
\end{gather}
Substitute (\ref{yvH7}) into (\ref{yvH6}) to see that $R^2=I_2+2
\pi J \b^* \b$, i.e., we can assume (\ref{yv2'}).

Also we can set
\[
A=(B-x I_n)^{-1}, \qquad B= \mathrm{diag} \{ b_1, b_2, \ldots ,
b_n \}.
\]
From $\Pi_x=-i A \Pi J H$  and (\ref{yv2}) we get
\begin{gather} \label{yv3}
\Pi J \b^* =g= \{g_k \}_{k=1}^n \equiv \mathrm{const}.
\end{gather}
We also have
\begin{gather*} 
\frac{d}{dx}\Pi(x)  \b^* =-2 i (B-xI_n)^{-1}\Pi(x) J \b^*.
\end{gather*}
It follows that
\begin{gather} \label{yv4}
\Pi(x)  \b^* = 2\big(\{ i g_k \ln (b_k-x)\}_{k=1}^n +h \big),
\qquad h \equiv \mathrm{const}.
\end{gather}
Formulas (\ref{yv3}) and (\ref{yv4}) give us $\Pi$. We shall
assume that $b_k \not\in [0, \infty) $, and so $\Pi$ is
well-def\/ined on $[0, \infty)$. Taking into account (\ref{yvH5})
we also get
\begin{gather}
\Pi(x)J \Pi(x)^*= \frac{1}{2} \Pi(x)T^*J T \Pi(x)^*
\nonumber\\
 \label{yv4'}
\phantom{\Pi(x)J \Pi(x)^*}{}=\big(\{ i g_k \ln (b_k-x)\}_{k=1}^n
+h \big)g^*+g\big(\{ i g_k \ln (b_k-x)\}_{k=1}^n +h \big)^*.
\end{gather}
The matrix function $S$ is  easily derived from the identity
$AS-SA^*= i \Pi J \Pi^*$.

Finally, in view of (\ref{yv1'}) and (\ref{yv3}) we get
\[
\b w_0(x)= \big(\b - i g^*S(x)^{-1}(B-x I_n) \Pi(x)\big) \wt U,
\]
which, taking into account (\ref{1.12}) and (\ref{yv2}), implies
\begin{gather} \label{yv5}
\wt H(x)= \wt U^* \big( \b - i g^*S(x)^{-1}(B-x I_n) \Pi(x)
\big)^*
 \big( \b - i g^*S(x)^{-1}(B-x I_n) \Pi(x) \big)\wt U.
\end{gather}
From (\ref{1.14}) it follows that $v(x,s)$ is $J$-unitary and we
rewrite (\ref{1.28'}) as
\begin{gather*} 
 \wt R(s)=I_2+ v(0,s) \big(R(s)-I_2 \big)J v(0,s)^*J.
\end{gather*}
Now by (\ref{1.7}), (\ref{1.10}), (\ref{yv2'}), and (\ref{yv3}) we
get
\begin{gather}
\wt R(s)=I_2+\pi J w_0(0)^* \big(\b^* - is
\Pi(0)^*S(0)^{-1}B(sI_n-B)^{-1}g \big)
\nonumber\\
 \label{yv6}
\phantom{\wt R(s)=}{}\times\big(\b^* - is
\Pi(0)^*S(0)^{-1}B(sI_n-B)^{-1}g \big)^* w_0(0).
\end{gather}
Thus matrix functions $\wt H(x)$ and $\wt R(s)$ are given by
formulas (\ref{yv5}) and (\ref{yv6}), respectively.

\begin{example}
Consider the simplest case $n=1$. Put $b_1=b$ and assume $b
\not\in \BR$. Rewrite (\ref{yv4'}) as
\begin{gather*} 
\Pi(x)J \Pi(x)^*=i |g|^2 \big( \ln (b-x)- \ov{\ln (b-x)} \big)+ h
\ov{g}+g \ov{h}.
\end{gather*}
Here $\ov{g}$ is the complex number conjugated to $g$. Hence, in
view of (\ref{1.6}) we get
\begin{gather*} 
S(x)= \big( |g|^2 \big( \ln (b-x)- \ov{\ln (b-x)} \big) -i h
\ov{g}-i g \ov{h} \big) \frac{(b-x)(\ov{b}-x)}{b- \ov{b}}.
\end{gather*}
Put now $h=0$ to derive
\begin{gather} \label{e3}
S(x)= \frac{2i}{b- \ov{b}} |g|^2 \big( \arg (b-x) \big)
(b-x)(\ov{b}-x) \not= 0.
\end{gather}
Rewrite (\ref{yv6}) as
\begin{gather} \label{e4}
\wt R(s)=I_2+ \pi J U^*r(s)^*r(s)U,
\end{gather}
where $U=w_0(0)$, $r(s)=\b+i \, \ov{b} \, \ov{g} \big( \ov{S(0)}
\big)^{-1}s(s- \ov{b})^{-1} \Pi(0)$. By (\ref{yvH5}), (\ref{yv3}),
and (\ref{yv4}) we have
\begin{gather} \label{e5}
\Pi(x)=\frac{1}{2}\Pi(x)T^*J T J=\frac{1}{2}g\Big[i\big(1+2 \ln
(b-x)  \big) \quad 1-2 \ln (b-x) \Big].
\end{gather}
Formulas (\ref{e3}) and (\ref{e5}) imply
\begin{gather} \label{e6}
r(s)=\b+ \frac{b- \ov{b}}{4b \arg \, b} \frac{s}{s- \ov{b}}
\Big[i\big(1+2 \ln \, b  \big) \quad 1-2 \ln \, b \Big].
\end{gather}
Equalities (\ref{e4}) and (\ref{e6}) def\/ine $\wt R(s)$
determined by the parameters $b$ and $g$ and $J$-unitary matrix
$U$. According to (\ref{1.7'}), (\ref{yv1'}), and (\ref{yv5}) the
corresponding matrix function $\wt H(x)$ is of the form
\begin{gather*} 
\wt H(x)= \wt U^*h(x)^*h(x) \wt U,
\end{gather*}
where $h(x)= \b - i \, \ov{g} \,(b-x) S(x)^{-1} \Pi(x)$, and
\begin{gather*} 
\wt U=Jw_A(0, \infty)^*J U=\left( I_2+ \frac{i \,
\ov{b}}{\ov{S(0)}}J \Pi(0)^* \Pi(0)\right) U.
\end{gather*}
Finally, using (\ref{e3}) and (\ref{e5}) rewrite $h$ in the
explicit form:
\begin{gather*} 
h(x)=\b-  \frac{b- \ov{b}}{4(\ov{b}-x) \arg \, (b-x)}
\Big[i\big(1+2 \ln (b-x)  \big) \quad 1-2 \ln (b-x) \Big].
\end{gather*}
\end{example}

\section{Inverse problem: explicit
solutions}\label{sec3}
 In view of (\ref{1.7'}) it is immediate
that formula (\ref{yv6}) can be written in the form (\ref{e4}):
$\wt R(s)=I_2+ \pi J U^*r(s)^*r(s)U$, where vector function
\begin{gather} \label{3.0}
r(s)=\b J w_A(0,s)^*J =\b + is g^*(sI_n-B^*)^{-1}B^*
S(0)^{-1}\Pi(0)
\end{gather}
 is
rational and satisf\/ies the following properties
\begin{gather} \label{3.1}
r(s)=[r_1(s) \quad r_2(s)] \in \BC^2, \qquad r(s)Jr(s)^*=0, \qquad
r(0)=\b.
\end{gather}
Here $J$ is def\/ined in (\ref{yv2}). Introduce matrices $K$ and
$j$:
\begin{gather} \label{3.2}
K:= \frac{1}{\sqrt{2}} \left[\begin{array}{lr}1 & -1
\\ 1 &
1\end{array} \right], \qquad j:=\left[\begin{array}{lr}1 & 0 \\ 0
& -1\end{array} \right], \qquad K^*=K^{-1}, \qquad K j K^*=J.
\end{gather}
Consider function
\begin{gather} \label{3.3}
u(s)=\big(\ov{r_1(s^{-1})}+\ov{r_2(s^{-1})}\big)
\big(\ov{r_1(s^{-1})}-\ov{r_2(s^{-1})}\big)^{-1}.
\end{gather}
From (\ref{3.1}) and (\ref{3.2}) we get $r K j K^* r^*=0$, and so
\begin{gather} \label{3.4}
|u| \equiv 1, \qquad u(\infty)=(1-i)/(1+i).
\end{gather}
Rational function $u$ satisfying (\ref{3.4}) admits \cite{SaL1} a
so called minimal realization
\begin{gather} \label{3.5}
u(s)=c^2 \big(1 + i \t^*S_0^{-1}(sI_n - \a)^{-1} \t \big), \qquad
c=(1-i)/ \sqrt{2},
\end{gather}
where
\begin{gather} \label{3.6}
\a S_0 - S_0 \a^*=i \t \t^*.
\end{gather}
Using (\ref{3.5}) one recovers $\wt H$ (explicitly, though not
necessarily uniquely) from the given function~$u$.

\begin{theorem}
Let $J$-unitary matrix $U$ and rational function $u$ satisfying
\eqref{3.4} be given. Consider realization \eqref{3.5} and choose
vectors $\t_1$ and $\t_2$ so that
\begin{gather} \label{3.7}
c \,  \t_1 + \ov{c} \, \t_2 = \t, \qquad \det \big( \a-i \, c \,
\t \t_2^* S_0^{-1} \big) \not=0.
\end{gather}
Supposing relations \eqref{3.7} are valid, put
\begin{gather} \label{3.8}
A(0):=\a-i \, c \, \t \t_2^* S(0)^{-1}, \qquad S(0):=S_0, \qquad
\Pi(0):= \L K^*, \qquad \L:=[\t_1 \quad \t_2].
\end{gather}
Introduce now matrix-functions $S$ and $\Pi$ by \eqref{3.8} and
equations
\begin{gather} \label{3.9}
S_x=g g^*-(AS+SA^*), \qquad \Pi(x)J \b^*=g, \qquad \Pi_x(x)
\b^*=-2i(B-xI_n)^{-1}g,
\end{gather}
where
\begin{gather} \label{3.10}
B=A(0)^{-1}, \qquad A(x)=(B-x I_n)^{-1}, \qquad g= \t.
\end{gather}
Then on the intervals $[0, \, l]$, where $\det(B-xI_n) \not=0$ and
$\det \, S(x) \not= 0$, matrix function $\wt R$ is well-defined by
\eqref{e4}, \eqref{3.0}, \eqref{3.8}, and \eqref{3.10}. Moreover,
equality \eqref{3.3} is true and $\wt H$ corresponding to $\wt R$
is given by \eqref{yv5}, \eqref{3.8}--\eqref{3.10}, and the second
relation in \eqref{yv1'}.
\end{theorem}

\begin{proof}
First notice that  equations  (\ref{3.6}) and (\ref{3.8}) and the
f\/irst relation in (\ref{3.7}) imply identities
\begin{gather*} 
A(0)S(0)-S(0)A(0)^*=i \L j \L^*=i \Pi(0) J \Pi(0)^*.
\end{gather*}
The correspondence between $\wt R$ and $\wt H$ follows now from
the results of Section \ref{sec2}. It remains to prove
(\ref{3.3}). For this purpose notice that by (\ref{1.7}),
(\ref{3.0}), (\ref{3.2}), and (\ref{3.8}) we have
\begin{gather} \label{3.12}
jK^*r(s^{-1})^*=W(s)jK^* \b^*=
W(s)\left[\begin{array}{c}c \\
\ov{c} \end{array} \right],
\end{gather}
where $2 \times 2$ matrix function $W$ is of the form
\begin{gather} \label{3.13}
W(s)=\{ W_{kj}(s) \}_{k,j=1}^2=I_2-ij \L^* S(0)^{-1}\big(A(0)-sI_n
\big)^{-1} \L.
\end{gather}
In view of (\ref{3.12}) we get
\begin{gather} \label{3.14}
\frac{\ov{r_1(s^{-1})}+\ov{r_2(s^{-1})}}{
\ov{r_1(s^{-1})}-\ov{r_2(s^{-1})}}=\frac{cW_{11}(s)+
\ov{c}W_{12}(s)}{cW_{21}(s)+ \ov{c}W_{22}(s)}.
\end{gather}
From the f\/irst relation in (\ref{3.7}) and (\ref{3.13}) follow
the representations
\begin{gather} \label{3.15}
cW_{11}(s)+ \ov{c}W_{12}(s)=c-i \t_1^* S(0)^{-1}\big(A(0)-sI_n
\big)^{-1}\t,
\\
 \label{3.16}
cW_{21}(s)+ \ov{c}W_{22}(s)=\ov{c}+i \t_2^*
S(0)^{-1}\big(A(0)-sI_n \big)^{-1}\t.
\end{gather}
Using  system theory results on the realization of the inverse
matrix function, from the f\/irst relation in (\ref{3.8}) and
(\ref{3.16}) we derive
\begin{gather} \label{3.17}
\big(cW_{21}(s)+ \ov{c}W_{22}(s)\big)^{-1}=c \big(1-i \, c \,
\t_2^* S(0)^{-1}\big(\a-sI_n \big)^{-1}\t \big).
\end{gather}
Finally, taking into account that $i(A(0)-\a)= c \, \t \t_2^*
S(0)^{-1}$ from (\ref{3.15}) and (\ref{3.17}) we get
\begin{gather} \label{3.18}
\frac{cW_{11}(s)+ \ov{c}W_{12}(s)}{cW_{21}(s)+ \ov{c}W_{22}(s)}=
c^2 \big(1 + i \t^*S_0^{-1}(sI_n - \a)^{-1} \t \big).
\end{gather}
Equalities (\ref{3.5}), (\ref{3.14}), and (\ref{3.18}) imply
(\ref{3.3}).
\end{proof}

Compare  the result with the recovery of the so called
pseudo-exponential potentials (see~\cite{FKS, GKS6} and references
therein).
\begin{remark} \label{Rk3.2}
As $|u|=1$, so matrix $\a$ in the realization (\ref{3.5}) is
invertible. Therefore we can choose  $\t_2$ satisfying the f\/irst
relation in (\ref{3.7}) and suf\/f\/iciently small for $\a-i \, c
\, \t \t_2^* S_0^{-1}$ to be invertible too.
\end{remark}

\section{Summary}\label{Sum}

The f\/irst new result in this paper is the construction of the
B\"acklund--Darboux transformation for the non-isospectral
canonical system (\ref{1.1}), which is important both in
prediction theory and random matrices theory. The GBDT-version of
the B\"acklund--Darboux transformation, constructed in  Theorem
\ref{TmBDT}, is more general than iterated BDT and admits
parameter matrix $A$ with an arbitrary Jordan structure. (For the
applications of GBDT to non-isospectral integrable systems see
\cite{SaA2}.)

In Section~3, we apply GBDT to the initial system with $H \equiv
\mathrm{const}$ to obtain a family of  explicit solutions of
system (\ref{1.1}) and of the corresponding Riemann--Hilbert
problem (\ref{1.24}). In particular, we construct the transformed
Hamiltonians $\wt H$ and the transformed jump functions~$\wt R^2$
(see formula (\ref{yv5}) for $\wt H$ and formula (\ref{yv6})   for
$\wt R$). The subcase from Example~1 is treated in greater detail.
The interesting case of non-diagonal matrix $A$ and applications
to prediction theory will follow elsewhere.

Finally, in Section~4, using the methods of system theory, we
recover  $\wt H$ and
 $\wt R$ from a partial information on $\wt R$ similar
to the way, in which the Dirac system is recovered explicitly from
its Weyl function in~\cite{GKS6}.

\subsection*{Acknowledgements}
The work was supported by the Austrian Science Fund (FWF) under
Grant  no.~Y330.

\pdfbookmark[1]{References}{ref}
\LastPageEnding


\begin{thebibliography}{99}

\footnotesize\itemsep=0pt

\bibitem{C}
Cieslinski J., An ef\/fective method to compute $N$-fold Darboux
matrix and $N$-soliton surfaces, {\it J. Math. Phys.} {\bf 32}
(1991), 2395--2399.

\bibitem{Dei0}
Deift P., Applications of a commutation formula, {\it Duke Math.
J.} {\bf 45} (1978),   267--310.

\bibitem{Dei00}
Deift P., Its A., Zhou X., A Riemann--Hilbert approach to
asymptotic problems arising in the theory of random matrix models,
and also in the theory of integrable statistical mechanics, {\it
Ann. of Math. (2)} {\bf 146} (1997), 149--235.

\bibitem{Dei}
Deift P.A., Orthogonal polynomials and random matrices: a
Riemann--Hilbert approach, in Courant Lecture Notes in
Mathematics, Vol.~3, AMS, Providence, RI, 1999.

\bibitem{FKS}
Fritzsche B., Kirstein B., Sakhnovich A.L., Completion problems
and scattering problems for Dirac type dif\/ferential equations
with singularities,  {\it J. Math. Anal. Appl.} {\bf 317} (2006),
 510--525, \href{http://arxiv.org/abs/math.SP/0409424}{math.SP/0409424}.

\bibitem{Ge}
 Gesztesy F., A complete spectral characterization of
the double commutation method, {\it J. Funct. Anal.} {\bf 117}
(1993),   401--446.

\bibitem{GKS6}
Gohberg I., Kaashoek M.A.,  Sakhnovich A.L., Scattering problems
for a canonical system with a pseudo-exponential potential,  {\it
Asymptotic Analysis} {\bf 29} (2002), 1--38.

\bibitem{Gu}
Gu C.,  Hu H.,  Zhou Z., Darboux transformations in integrable
systems, {\it Math. Phys. Stud.}, Vol.~26,  Springer, Dordrecht,
2005.

\bibitem{KPR}
 Kuznetsov V.B., Petrera M.,  Ragnisco O., Separation
of variables and B\"acklund transformations for the symmetric
Lagrange top,  {\it J. Phys. A: Math. Gen.} {\bf 37} (2004),
8495--8512,
\href{http://arxiv.org/abs/nlin.SI/0403028}{nlin.SI/0403028}.

\bibitem{KSS}
Kuznetsov V.B., Salerno M., Sklyanin E.K., Quantum B\"acklund
transformation for the integrable DST model, {\it J. Phys. A:
Math. Gen.} {\bf 33} (2000), 171--189,
\href{http://arxiv.org/abs/solv-int/9908002}{solv-int/9908002}.

\bibitem{Mar}
Marchenko V.A.,  Nonlinear equations and operator algebras, Reidel
Publishing Co., Dordrecht, 1988.

\bibitem{MS}
Matveev V.B., Salle M.A.,  Darboux transformations and solitons,
Springer, Berlin, 1991.

\bibitem{Mi}
Miura R.  (Editor),  B\"acklund  transformations, {\it Lecture
Notes in Math.}, Vol.~515, Springer,  Berlin, 1976.

\bibitem{SaA2}
Sakhnovich A.L., Iterated B\"acklund--Darboux transformation and
transfer matrix-function (nonisospectral case), {\it Chaos
Solitons Fractals} {\bf 7} (1996), 1251--1259.

\bibitem{SaA3}
Sakhnovich A.L., Iterated B\"acklund--Darboux transform for
canonical systems, {\it J. Funct. Anal.} {\bf 144} (1997),
359--370.

\bibitem{SaL5}
Sakhnovich L.A., Operators, similar to unitary operators, with
absolutely continuous spectrum, {\it Funct. Anal. Appl.} {\bf 2}
(1968), 48--60.

\bibitem{SaL1}
Sakhnovich L.A., On  the  factorization  of  the transfer matrix
function,  {\it Sov. Math. Dokl.} {\bf 17} (1976), 203--207.

\bibitem{SaL2}
Sakhnovich L.A., Factorisation  problems  and operator identities,
{\it Russian
 Math. Surv.} {\bf 41} (1986), 1--64.

\bibitem{SaL3}
Sakhnovich L.A., Spectral theory of canonical dif\/ferential
systems. Method of operator identities, {\it Oper. Theory Adv.
Appl.}, Vol.~107, Birkh\"auser, Basel~-- Boston, 1999.

\bibitem{SaL4}
Sakhnovich L.A.,  Integrable operators and canonical
dif\/ferential systems, {\it Math. Nachr.} {\bf 280} (2007),
205--220,
\href{http://arxiv.org/abs/math.FA/0403490}{math.FA/0403490}.

\bibitem{T}
Teschl G., Jacobi operators and completely integrable nonlinear
lattices, {\it Mathematical Surveys and Monographs},  Vol.~72,
 AMS, Providence, RI, 2000.

\bibitem{W}
Wiener N., Extrapolation, interpolation, and smoothing of
stationary time series, Chapman and Hall  Ltd., London, 1949.

\bibitem{ZM}
Zakharov V.E., Mikhailov A.V.,  On the integrability of classical
spinor models in two-dimensional space-time, {\it Comm. Math.
Phys.} {\bf 74} (1980), 21--40.

\end{thebibliography}
\end{document}